\documentclass[rnote]{aa} % for the research notes

\usepackage{graphicx}
\usepackage{longtable}
\usepackage{lscape}
\usepackage{txfonts}
\usepackage{amsfonts}
\usepackage{rotating}
\usepackage{natbib}
\RequirePackage{amssymb}
\RequirePackage{latexsym}

\bibpunct{(}{)}{;}{a}{}{,} % to follow the A&A style
\begin{document}
\title{On the binarity of the classical Cepheid X~Sgr\\
from interferometric observations\thanks{Based on observations collected at the European Southern Observatory, Paranal, Chile}}

   \author{G.~Li~Causi
   		\inst{1,2},
   		S.~Antoniucci\inst{1},
   		G.~Bono\inst{3},
   		S.~Pedicelli\inst{1},
   		D.~Lorenzetti\inst{1},
		T.~Giannini\inst{1},
   		B.~Nisini\inst{1}}

	\institute{INAF-Osservatorio Astronomico di Roma, Via Frascati 33, I-00040 Monteporzio Catone (RM) - Italia	\and	INAF-Istituto di Astrofisica e Planetologia Spaziale, Via Fosso del Cavaliere 100, Roma - Italia	\and	Universit\`a degli Studi di Roma `Tor Vergata', via della Ricerca Scientifica 1, I-00133 Roma - Italia}

	\offprints{Gianluca Li Causi, \email{gianluca.licausi@inaf.it}}

%   \date{Received September 15, 1996; accepted March 16, 1997}

  \abstract
  {Optical-infrared interferometry can provide direct geometrical measurements of the radii of Cepheids and/or reveal unknown binary companions of these stars. Such information is of great importance for a proper calibration of Period-Luminosity relations and for determining binary fraction among Cepheids.}
   {We observed the Cepheid X~Sgr with VLTI/AMBER in order to confirm or disprove the presence of the hypothesized binary companion and to directly measure the mean stellar radius, possibly detecting its variation along the pulsation cycle.}
   {From AMBER observations in MR mode we performed a binary model fitting on the closure phase and a limb-darkened model fitting on the visibility.}
   {Our analysis indicates the presence of a point-like companion at a separation of 10.7~mas and 5.6~mag$_K$ fainter than the primary, whose flux and position are sharply constrained by the data. The radius pulsation is not detected, whereas the average limb-darkened diameter results to be 1.48$\pm$0.08~mas, corresponding to 53$\pm$3~R$_{\odot}$. at a distance of 333.3~pc.}
   {}


	\authorrunning{Li Causi et. al}
	\titlerunning{AMBER binarity of X~Sgr}
   \maketitle

%
%________________________________________________________________

\section{Introduction}

Classical Cepheids are fundamental astrophysical objects, being the most popular
primary distance indicators in the nearby Universe.
These pulsating stars obey well defined Period-Luminosity (PL) relations, so that it is possible to derive their distance moduli by measuring periods and apparent magnitudes.

Cepheids can be used to trace the evolutionary properties of Helium burning intermediate-mass stars, as they are crossing the instability strip along the so-called ``blue loops'' in the HR diagram.
However, the comparison between pulsation and evolutionary masses 
disclosed the problem known as \textit{``Cepheid mass discrepancy''}
\citep{COX_1980}, due to the pulsation masses resulting systematically smaller
than the evolutionary ones (see e.g. \citet{BONO_2001}; \citet{Beaulieu_2001}; \citet{CAPUTO_2005}; \citet{KELLER_2006}).

Recently two double eclipse binary Cepheids have been identified in the
Large Magellanic Cloud (\citet{PIETRZYNSKI_2011}, \citet{PIETRZYNSKI_2011}), providing the first measurements
of the dynamical mass of classical Cepheids with 1-3\% accuracy, whose values support
the pulsation masses, or the Mass-Luminosity relations based on evolutionary models
that account for mild convective core overshooting (\citet{CASSISI_SALARIS_2011}; \citet{PRADA_MORONI_2012}).

For non-eclipsing binaries, direct orbital determination of the companion by means of
optical interferometry would provide dynamical mass determinations able to disentangle
the mass discrepancy problem.

The binary fraction of the Cepheids has an impact on the Cepheid distance 
scale in the optical bands, since the companions are typically 
main sequence (MS) stars. The binarity analysis has been applied to 
a limited sample \citep{SZABADOS_2003}, but current estimates
indicate that the binary fraction among MS B-type stars is of the order
of 60--70\% (\citet{BROTT_2011}; \citet{CHINI_2012}).

In order to improve the current observational scenario we have observed 
the Cepheid X~Sgr with the VLTI. This particular target was selected because:
{\em i)} its trigonometric parallax has been recently measured by 
\citet{BENEDICT};
{\em ii)} accurate radial velocity curves are available
(\citet{Mathias_2006}; \citet{STORM_2011}), and 
{\em iii)} it was suggested by \citet{SZABADOS_1990}, on the basis of temporal 
changes in the $\gamma$-velocity, that X~Sgr might be a binary, although this finding has always been controversial (\citet{Mathias_2006}; \citet{BAADE-WESSELINK}).

%%____END OF "Introduction" SECTION_____

%%%%%%%%%%%%%%%%%%%%%%%%%%%%%%%%%%%%%%%%%%%%%%%%%%%%%%%%%%%%%%%%%%%%%%%%%%%%%%%%%%%%%%%%
\section{Observations and data reduction}

\subsection{Data acquisition}

We observed X~Sgr (HD161592, R.A.~17:47:33.62, Dec.~-27:49:50.83, J2000) 
using VLTI/AMBER \citep{Petrov_AMBER} with the Auxiliary Telescopes.
We used medium spectral resolution in the $K$-band (MR, R$\sim$1500, 2.1$\mu$m)
and the fringe tracker FINITO \citep{FINITO} with a frame
exposure time of 1~s. Observations were carried out with a maximum projected baseline length 
of $127.8$~m, corresponding to $3.1$~mas spatial resolution,
with the atmospheric seeing varied in the $0.7"\div1.0"$ range.

%\begin{center}
	\begin{table*}
    \centering
	\caption{Log of the interferometric observations, with details on the 
SCIence-CALibrators sequence used and the optimal frame selection adopted for $V^2$ 
and $\phi$. Magnitude and diameters of the two calibrators employed are shown.}
	\label{Table_Observations_Log}      
	\begin{tabular}{clccc}		%c for center, l for left
	\hline\hline       
	Date & Target sequence & Baselines & $\%$ of frames for $V^2$ & $\%$ of frames for $\phi$ \\
	\hline                    
			2011-04-24 & CAL$_2$-SCI-CAL$_2$-CAL$_1$-SCI-CAL$_1$ 				& D0-I1-G1 & 10$\%$ & 100$\%$ \\
			2011-05-03 & CAL$_2$-SCI-CAL$_1$-SCI-CAL$_1$-CAL$_2$-SCI-CAL$_2$ 	& D0-A1-C1 & 30$\%$ & 100$\%$ \\
			2011-05-07 & CAL$_1$-SCI-CAL$_1$-SCI-CAL$_1$						& D0-A1-C1 & 10$\%$ & 100$\%$ \\
			2011-06-11 & CAL$_1$-SCI-CAL$_1$ 									& K0-A1-I1 & 50$\%$ & 100$\%$ \\
			2011-06-25 & CAL$_1$-SCI-CAL$_1$-SCI 								& K0-A1-I1 & 20$\%$ & 100$\%$ \\
			2012-03-31 & CAL$_2$-SCI-CAL$_1$-SCI-CAL$_1$						& K0-A1-G1 & 20$\%$ & 100$\%$ \\
	\\
	\hline\hline
	ID	& Name & mag$_K^a$ & $\theta_{UD}^a$ [mas] \\
	\hline                    
		CAL$_1$ & HIP88839 & 2.070 & 1.881 $\pm$ 0.133 \\
		CAL$_2$ & HD157919 & 3.203 & 0.947 $\pm$ 0.066 \\
	\hline                  
	\end{tabular}
	\begin{list}{}{}
		\item[$^a$] Data from JMMC Stellar Diameters Catalogue - JSDC \citep{Lafrasse_2010}
	\end{list}	
	\end{table*}
%\end{center}
	
For the sake of clarity, we distinguish here a Run~A (six nights from 2011-04-24 to 2011-06-25)
and a Run~B (2012-03-31 night only, see Table~\ref{Table_Observations_Log} and Fig.~\ref{Fig_V2}).
For each run we used a CAL-SCI-CAL observation scheme (Table~\ref{Table_Observations_Log}) to better sample the variable instrument response during the night.

%---- END OF "Data Aquisition" SUBSECTION

%%%%%%%%%%%%%%%%%%%%%%%%%%%%%%%%%%%%%%%%%%%%%%%%%%%%%%%%%%%%%%%%%%%%%%%%%%%%%
\subsection{Data reduction}
\label{Data Reduction}

We used the \textit{amdlib v3}\footnote{Available at http://www.jmmc.fr/amberdrs} 
(\citet{Tatulli_AMBER}, \citet{CHELLI_2009}) reduction package in conjunction with custom IDL 
scripts for optimizing the fringe selection, computing averages, and calibrating the visibilities.

Each AMBER file contains many individual frames, which must be quality-selected
to reject tracking losses in both FINITO and telescope guiding.
Such events are associated to large values in the optical path difference 
(OPD) and cause a poor signal to noise ratio (SNR) of the fringes.
We adopted the \textit{amdlib} ``PISTON+SNR'' selection criteria, limiting
the OPD variation below 8$\mu$m and keeping only the frames with the highest fringe SNR.
The adopted percentage of frames, in the range $10\div50\%$ for
visibility $V^2$ and $100\%$ for closure phase $\phi$, has been optimized
for each night separately (see Table~\ref{Table_Observations_Log}).

Hence, we computed, for each file, the average values of $V^2$ and $\phi$, weighted by
the error estimate provided by amdlib; the same was done for the calibrators.

Finally, a night transfer function (TF) was computed using the calibrators
listed in Table~\ref{Table_Observations_Log}.

%---- END OF "Data Reduction" SUBSECTION

%%____END OF "Observation and Reduction" SECTION_____

%%%%%%%%%%%%%%%%%%%%%%%%%%%%%%%%%%%%%%%%%%%%%%%%%%%%%%%%%%%%%%%%%%%%%%%%
\section{Results and Discussion}

\subsection{Visibilities and closure phases}
\label{Visibilities and Closure Phases}

\begin{figure*}
   \centering
   \includegraphics[height=6cm]{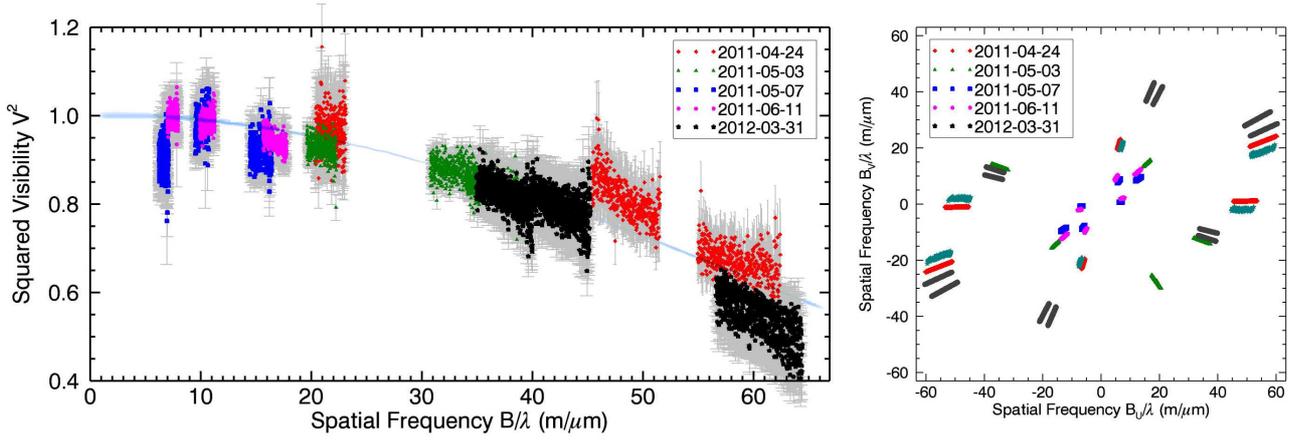}
   \caption{Left: calibrated $V^2$ versus spatial frequency B/$\lambda$ 
for the entire set of observations, superimposed to the average UD model (continuous line). Right: spatial frequencies (\textit{u,v}-plane) coverage. 
Different colors and symbols refer to different observing nights 
(see legends). Note: the 2011-06-25 night is not plotted in the left panel, due to a very
unstable TF for $V^2$.}
   \label{Fig_V2}
   \end{figure*}

\begin{figure*}
   \centering
   \includegraphics[height=6cm]{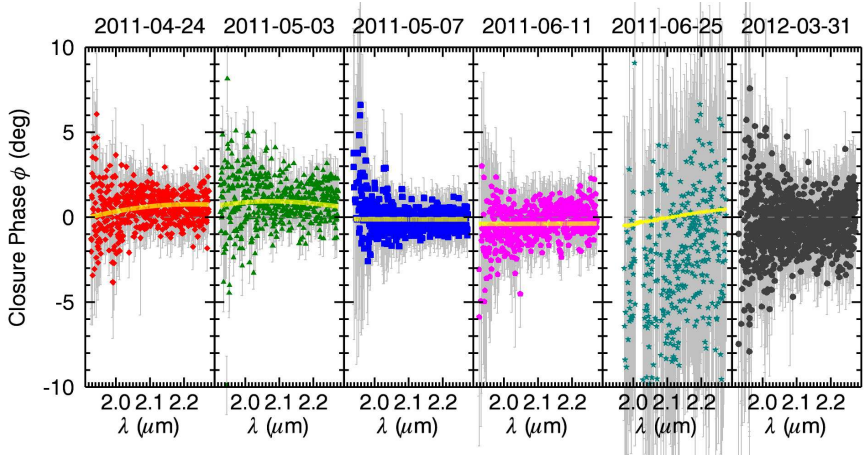}
   \caption{Calibrated $\phi$ versus $\lambda$ for the entire set 
of observations. Colors and symbols are the same as in 
Fig.~\ref{Fig_V2}. The yellow line shows the binary model 
described in Sec.~\ref{Model fitting} providing the best fit to 
the Run~A observations. The dates reported above the panels refer to 
the observing nights.}
   \label{Fig_CP}
   \end{figure*}

Fig.~\ref{Fig_V2} shows the calibrated $V^2$ as a function of the spatial 
frequency B/$\lambda$, while Fig.~\ref{Fig_CP} displays the calibrated $\phi$ 
versus wavelength $\lambda$.

%\begin{center}
%\end{center}

%\begin{center}
%\end{center}

We see that the source is resolved, because $V^2$ is less than unity, and that
significant deviations from a pure Uniform Disk (UD) are evident.
This could indicate a possible detection of the Cepheid pulsation, or might suggest a more complex 
morphology, like that of a binary.
In any case, in the $\phi$ plot we only see small deviations from zero, which is an indication  
that the companion, if present, should be faint and close to our detection limit.

Both plots, in particular the $V^2$, show that the error bars are clearly larger than the dispersion of the data points,
indicating a correlated noise caused by the low-frequency variations (on time-scales of minutes) of the TF during the night.

%---- END OF "Visibilities and Closure Phases" SUBSECTION

%%%%%%%%%%%%%%%%%%%%%%%%%%%%%%%%%%%%%%%%%%%%%%%%%%%%%%%%%%%%%%%%%%%%%%%%
\subsection{Model fitting}
\label{Model fitting}

\subsubsection{Binary companion}
\label{Binary companion}

\citet{SZABADOS_1990} suggested, on the basis of 
changes in the $\gamma$-velocity, that X~Sgr might be a binary system with a
period of $\sim$507~days.
The complex nature of X~Sgr was further supported by 
\citet{Saselov_Lester_1990} who spectrally found that line asymmetries and
line-doubling were well beyond the typical Cepheids behaviour; in
particular, the two absorption components showed different Doppler shifts, but they
did not display secular velocity changes compared with the 
$\gamma$-velocity. This complex dynamical structure was also supported by other
high-resolution (R$\sim$120,000) optical spectra by \citet{Mathias_2006}. 
They hypothesized a binary, or even triple, nature of X~Sgr.
The data available in the literature 
were analyzed by \citet{FEAST_XSGR_DATA} and by 
\citet{BAADE-WESSELINK} who also found evidence of binarity, these latter
suggesting a slightly longer orbital period ($\sim$570~days).

In order to check the companion hypothesis, we performed
a binary model fit to the closure phase $\phi$. We decided to 
use only this observable, for its low sensitivity on
the unknown radius variation, for which we adopted a crude
UD estimate of $r_{UD}=0.74$~mas, provided by the $V^2$ values at the 
largest baselines, namely the data from nights 
2011-04-24, 2011-05-03 and 2012-03-31.

We started with Run~A alone assuming no temporal variation in the binary configuration within
the two months, which is a plausible assumption, considering the mentioned
orbital period estimates. Therefore we fit the Run~A data all together
in order to increase accuracy. Run~B was not included here, since it was acquired 10~months later.

We adopted a grid search on a wide parameter space in 
$\Delta x, \Delta y, f$ and $r_2$, where $\Delta x, \Delta y$ are 
the R.A. and Dec. offsets of the companion (these are 
searched within a radius of 100~mas around the primary, see \citet{ABSIL_PIONIER}), 
$f$ is the flux ratio of the companion to the primary, 
and $r_2$ is the companion UD radius.

\begin{figure*}
   \centering
   \includegraphics[height=6cm]{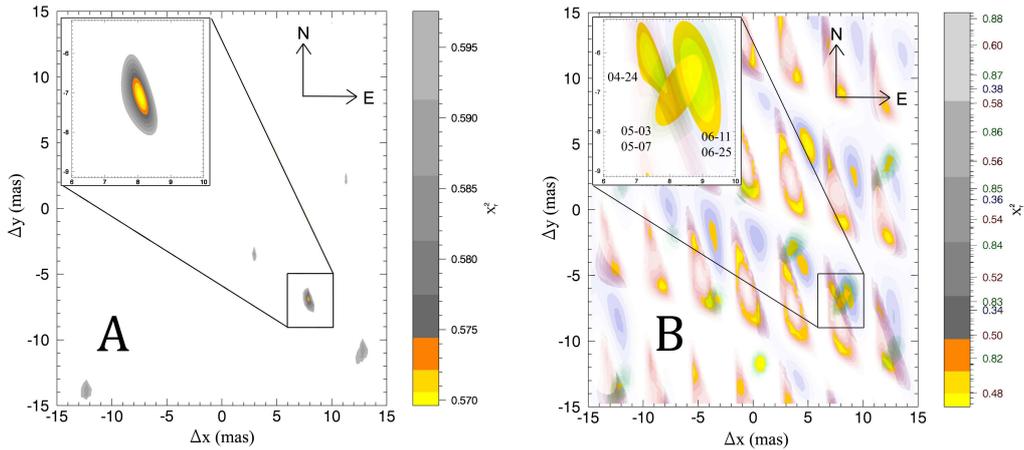}
   \caption{Portion of the reduced $\chi^2$ map on the 
R.A. and Dec. offsets obtained by A) fitting the $\phi$ of the 
Run~A data sets together with a point-like companion model at fixed flux 
ratio $f=6\cdot10^{-3}$, and B) for individual observing nights: red for 2011-04-24,
green for 2011-05-03/07, and blue for 2011-06-11/25.
The $1\div3\sigma$ confidence contours colours range from yellow to orange.}
   \label{Fig_Map_CP_all}
   \end{figure*}

We obtain the best fit result for a flux ratio of $f=(6\pm1)\cdot10^{-3}$
(corresponding to a $5.6\pm0.2$ magnitude difference), 
while $r_2$ tends to zero, thus confirming the presence of a faint and unresolved 
companion, whose position is constrained by a single and well 
defined $\chi^2$ minimum at 7.15$\pm0.04$~S, 8.00$\pm0.02$~E~mas from the primary
(see Fig~\ref{Fig_Map_CP_all}). The parameters of the companion are summarized in Table~\ref{Table_Binary_Parameters}.

The fact that the best fit reduced $\chi^2$ is significantly 
smaller than one is due to the large error bars associated 
to the TF, already described in Sec.~\ref{Visibilities and Closure Phases}.

%\begin{center}
%\end{center}

To double check our finding, 
we built a series of synthetic binary models with different 
flux ratios and the same error bars of the real data, from which
we derived a sensitivity limit of $f_{lim}=2\cdot10^{-3}$, i.e.
3 times lower than the measured flux ratio.

To refine this result we removed our assumption of fixed binary configuration,
in order to check if we could detect possible displacements among different nights.

Therefore, we repeated the $\phi$ fitting separately for each night 
of Run~A, to find out if the location of the $\chi^2$ minimum moves 
as a function of time (nights 2011-05-03/2011-05-07 and 2011-06-11/2011-06-25 
were fitted as single points).

The corresponding $\chi^2$ maps (Fig.~\ref{Fig_Map_CP_all}, panel~A) show several local minima within the 99$\%$ confidence interval, all of them having the same probability 
to be the true position of the companion. This shows us that a single baseline configuration
is not sufficient to unambiguously locate such a faint object.

However, each $\chi^2$ map shows a particular local minimum in the same position
found using the entire data set, and even more importantly the little displacements
of these minima are in the correct temporal sequence,
as disclosed by Fig.~\ref{Fig_Map_CP_all}, panel~B.

This empirical evidence further confirms our detection of the postulated companion of X~Sgr, and although current data do not allow us to constrain the orbit, we have a hint on the direction of its motion.

The binary fit for the single night of Run~B also show
the same kind of multiple minima $\chi^2$ map (not displayed), but with no minima close to 
the quoted position, as expected after a time interval of ten months.

%%\begin{center}
%   \begin{figure*}
%   \centering
%   \includegraphics[height=8cm]{Fig_Map_CP_each_night.eps}
%   \caption{Same as Fig.~\ref{Fig_Map_CP_all}, but for indivual 
%observing night: 2011-04-24, red; 2011-05-03/07, green; 
%2011-06-11/25, blue.}
%   \label{Fig_Map_CP_each_night}
%   \end{figure*}
%%\end{center}

%\begin{center}
	\begin{table}
%    \centering
	\caption{Parameters of X~Sgr companion, and LD diameters. Random errors (first ones) and
	systematics (second ones) are distinguished.}
	\label{Table_Binary_Parameters}      
	\begin{tabular}{cccc}		%c for center, l for left
	\hline\hline       
	$\Delta x$~[mas] & $\Delta y$~[mas] & $r_2$~[mas] & $f$ \\
	\hline                    
	8.00$\pm$0.02~E 	&	7.15$\pm$0.04~S	&	$<$0.5	& 	(6$\pm$1)$\cdot 10^{-3}$\\
	\\
	\hline\hline       
	Phase$^a$ & Date & $\theta_{LD}$ [mas] \\
	\hline
		0.83	&	2011-05-03	&	1.47$\pm$0.01$\pm$0.04	\\               
		0.54	&	2011-04-24	&	1.36$\pm$0.09$\pm$0.04	\\
		0.35	&	2012-03-31$^b$	&	1.59$\pm$0.08$\pm$0.04	\\
	\hline
	\end{tabular}
	\begin{list}{}{}
		\item[$^a$] The pulsation phase was estimated using the formula 
provided by \citet{SZABADOS_1989} for constraining the pulsation period and 
the epoch of maximum of the different data sets.
		\item[$^b$] For the 2012-03-31 night a pure LD model is adopted, given the undetermined binary parameters.
	\end{list}
	\end{table}
%\end{center}
	
%---- END OF "BINARY" SUBSUBSECTION

%%%%%%%%%%%%%%%%%%%%%%%%%%%%%%%%%%%%%%%%%%%%%%%%%%%%%%%%%%%%%%%%%%%%%%%%%%%%%%%%%%%
\subsection{Angular diameter}

To determine the mean Cepheid diameter, and possibly its variation, 
we fit the $V^2$ separately for each night by fixing the binary 
parameters (Table~\ref{Table_Binary_Parameters}) and varying only the 
primary radius $r_1$. Following \citet{KERVELLA_XSGR}, we adopted the  
Limb Darkening (LD) intensity profile by \citet{CLARET_LD_MODEL}. 

The radius $r_1$ only depends on the mean
value of each visibility spectrum ($<V^2(\lambda)>$), whose error bar we computed as
the quadratic sum of the weighted average error plus the correlated component due to the TF variation\footnote{where the latter was estimated, for each visibility spectrum, 
as the squared difference between the r.m.s. of the error bars and the r.m.s. of the 
residuals with respect to a smooth polynomial fitting.}.

%%\begin{center}
%   \begin{figure*}
%   \centering
%   \includegraphics[height=4.5cm]{Fig_Mean_Vis2_and_Physical_radius.eps}
%   \caption{Left: average visibility spectra $<V^2(\lambda)>$ showing the error
%bars described in the text. Right: comparison of the cepheid average radius 
%according to this work, to VINCI observations by \citet{KERVELLA_XSGR},
%and to IRSB estimates provided by \citet{STORM_2011} and by Groenewegen (private 
%communication): a significant fraction of current uncertainties 
%is introduced by the error on the trigonometric parallax; the open circle
%shows the radius estimate obtained with a simple LD model, neglecting the
%presence of the discovered companion.}
%   \label{Fig_Mean_Vis2}
%   \end{figure*}
%%\end{center}

%%\begin{center}
%   \begin{figure}
%   \centering
%   \includegraphics[height=4.5cm]{XSgr_Physical_Radius.eps}
%   \caption{Comparison of the cepheid average radius 
%according to this work, to VINCI observations by \citet{KERVELLA_XSGR},
%and to IRSB estimates provided by \citet{STORM_2011} and by Groenewegen (private 
%communication): \textbf{a significant fraction of current uncertainties 
%is introduced by the error on the trigonometric parallax}; the open circle
%shows the radius estimate obtained with a simple LD model, neglecting the
%presence of the discovered companion.}
%   \label{Fig_Physical_Radius}
%   \end{figure}
%%\end{center}

For the diameter determination we only took into account the nights 
2011-04-24, 2011-05-03 and 2012-03-31, which have sufficiently 
long baselines and good quality. The night of 2011-05-03 plays a key role in this context, 
since it shows a pure random error with a negligible correlated component.

Table~\ref{Table_Binary_Parameters} provides, in the bottom section, the best fit
LD diameters as a function of the Cepheid pulsation phase, computed following \citet{SZABADOS_1989}.

In the table we distinguish the random errors and 
the systematics coming from the uncertainties in the calibrator diameters
(Table~\ref{Table_Observations_Log}). Note that the latter are 
larger than the former only for our best data set, namely the night 
2011-05-03, showing the importance of using many different calibrators 
when a very accurate diameter measurement is required.

%\begin{center}
	\begin{table}
    \centering
	\caption{Recent X~Sgr radius estimates.}
	\label{Table_Radius_Comparison}
	\begin{tabular}{ccc}		%c for center, l for left
	\hline\hline       
	Article & Radius [R$_{\odot}$] \\
	\hline
	This paper	        	&	53 $\pm$ 3	\\
	This paper$^a$          &	55 $\pm$ 4	\\
	\citet{KERVELLA_XSGR}	&	53 $\pm$ 3	\\
	\citet{STORM_2011}$^b$	& 	50 $\pm$ 3	\\
	Groenewegen$^{b,c}$	    &	49 $\pm$ 3	\\
	\hline
	\end{tabular}
	\begin{list}{}{}
		\item[$^a$] Solution obtained by neglecting the presence of the companion.
		\item[$^c$] IRSB method.
		\item[$^c$] Private communication.
	\end{list}
	\end{table}
%\end{center}

The differences in the three diameter estimates, within $\sim 1~\sigma$ from each other,
do not allow us to positively detect the radius pulsation, and are probably due to the mentioned TF temporal variation.
The same limitation was also found by \citet{KERVELLA_XSGR} using interferometric data
collected with VINCI at VLTI.

Thus we only computed a mean physical radius of 53$\pm$3~R$_{\odot}$ by adopting the trigonometric parallax 
of 3.0$\pm$0.18~mas from the HST \citep{BENEDICT}.
In Table~\ref{Table_Radius_Comparison} we compared it  with recent estimates from
both optical interferometry \citep{KERVELLA_XSGR} and 
Infra Red Surface Brightness (IRSB, Groenewegen (personal communication) and \citet{STORM_2011}), having used the same distance for all of these measurements.
Finally, we also report the radius obtained by neglecting the binary companion, which has
a $\sim$3\% impact on the mean radius.

%____ END OF "Model Fitting" SECTION ______

%%%%%%%%%%%%%%%%%%%%%%%%%%%%%%%%%%%%%%%%%%%%%%%%%%%%%%%%%%%%%%%%%%%%%%%%%%%%%%%%%%%%%%%%%%%%%%
\section{Conclusions}

We found evidence of the binary nature of the Cepheid X~Sgr by analysing new interferometric data collected with AMBER at VLTI. 
The main conclusions of our work can be summarized as follows:

{\em i)} Our analysis of the closure phase indicates that both the flux and
the position of the companion can be constrained by our data sets.
We find that the angular separation is 10.7~mas and that the companion
is 5.6~mag fainter than the primary in the K-band.

{\em ii)} The data allow us to directly measure an average
Limb-Darkening diameter of 1.48$\pm$0.08~mas (i.e. a 
physical radius of 53$\pm$3~R$_{\odot}$) which agrees,  
at 1$\sigma$ level, with similar interferometric 
and IRSB mean radii for X~Sgr available in the literature.

{\em iii)} We find that MR AMBER closure phases obtained
from 3 baseline configurations have the sensitivity needed to 
detect companions around nearby 
classical Cepheids. This evidence becomes even more compelling 
if we take into account that current data were collected in medium
quality seeing conditions. 

{\em iv)} The average visibilities of the 2001-05-03 night show
that a non systematic error as
small as 1\% can be obtained with AMBER in MR mode in stable TF conditions.
This would be sufficient to clearly trace the differential diameter pulsation in future
measurements, provided that the same calibrator star is used for all the observations.

{\em v)} We find that, in the case of a binary Cepheid, the effect introduced by the companion on the primary radius measurement ($\sim$3\%) is comparable with the expected radius variation ($\sim$5 $\div$ 10\%).

%____ END OF "Conclusions" SECTION ______

\begin{acknowledgements}

We are grateful to the Nice Observatory (France), and in 
particular to F.~Millour, N.~Nardetto and P.~Cruzalebes for helpful 
discussions, and to M.A.T.~Gronewegen and J.~Storm for many useful 
insights concerning Cepheids and for providing us with their recent radius 
measurements of X~Sgr.
\end{acknowledgements}

\bibliographystyle{aa} % style aa.bst
\bibliography{Xsgr} % your references Yourfile.bib (references.bib scritto senza .bib)

\begin{thebibliography}{27}
\expandafter\ifx\csname natexlab\endcsname\relax\def\natexlab#1{#1}\fi

\bibitem[{{Absil} {et~al.}(2011){Absil}, {Le Bouquin}, {Berger}, {Lagrange},
  {Chauvin}, {Lazareff}, {Zins}, {Haguenauer}, {Jocou}, {Kern}, {Millan-Gabet},
  {Rochat}, \& {Traub}}]{ABSIL_PIONIER}
{Absil}, O., {Le Bouquin}, J.-B., {Berger}, J.-P., {et~al.} 2011, \aap, 535,
  A68

\bibitem[{{Beaulieu} {et~al.}(2001){Beaulieu}, {Buchler}, \&
  {Koll{\'a}th}}]{Beaulieu_2001}
{Beaulieu}, J.~P., {Buchler}, J.~R., \& {Koll{\'a}th}, Z. 2001, \aap, 373, 164

\bibitem[{{Benedict} {et~al.}(2007){Benedict}, {McArthur}, {Feast}, {Barnes},
  {Harrison}, {Patterson}, {Menzies}, {Bean}, \& {Freedman}}]{BENEDICT}
{Benedict}, G.~F., {McArthur}, B.~E., {Feast}, M.~W., {et~al.} 2007, \aj, 133,
  1810

\bibitem[{{Bono} {et~al.}(2001){Bono}, {Gieren}, {Marconi}, {Fouqu{\'e}}, \&
  {Caputo}}]{BONO_2001}
{Bono}, G., {Gieren}, W.~P., {Marconi}, M., {Fouqu{\'e}}, P., \& {Caputo}, F.
  2001, \apj, 563, 319

\bibitem[{{Brott} {et~al.}(2011){Brott}, {Evans}, {Hunter}, {de Koter},
  {Langer}, {Dufton}, {Cantiello}, {Trundle}, {Lennon}, {de Mink}, {Yoon}, \&
  {Anders}}]{BROTT_2011}
{Brott}, I., {Evans}, C.~J., {Hunter}, I., {et~al.} 2011, \aap, 530, A116

\bibitem[{{Caputo} {et~al.}(2005){Caputo}, {Bono}, {Fiorentino}, {Marconi}, \&
  {Musella}}]{CAPUTO_2005}
{Caputo}, F., {Bono}, G., {Fiorentino}, G., {Marconi}, M., \& {Musella}, I.
  2005, \apj, 629, 1021

\bibitem[{{Cassisi} \& {Salaris}(2011)}]{CASSISI_SALARIS_2011}
{Cassisi}, S. \& {Salaris}, M. 2011, \apjl, 728, L43

\bibitem[{{Chelli} {et~al.}(2009){Chelli}, {Utrera}, \& {Duvert}}]{CHELLI_2009}
{Chelli}, A., {Utrera}, O.~H., \& {Duvert}, G. 2009, \aap, 502, 705

\bibitem[{{Chini} {et~al.}(2012){Chini}, {Hoffmeister}, {Nasseri}, {Stahl}, \&
  {Zinnecker}}]{CHINI_2012}
{Chini}, R., {Hoffmeister}, V.~H., {Nasseri}, A., {Stahl}, O., \& {Zinnecker},
  H. 2012, \mnras, 424, 1925

\bibitem[{{Claret}(2000)}]{CLARET_LD_MODEL}
{Claret}, A. 2000, \aap, 363, 1081

\bibitem[{{Cox}(1980)}]{COX_1980}
{Cox}, A.~N. 1980, \araa, 18, 15

\bibitem[{{Feast} {et~al.}(2008){Feast}, {Laney}, {Kinman}, {van Leeuwen}, \&
  {Whitelock}}]{FEAST_XSGR_DATA}
{Feast}, M.~W., {Laney}, C.~D., {Kinman}, T.~D., {van Leeuwen}, F., \&
  {Whitelock}, P.~A. 2008, \mnras, 386, 2115

\bibitem[{{Gai} {et~al.}(2004){Gai}, {Menardi}, {Cesare}, {Bauvir}, {Bonino},
  {Corcione}, {Dimmler}, {Massone}, {Reynaud}, \& {Wallander}}]{FINITO}
{Gai}, M., {Menardi}, S., {Cesare}, S., {et~al.} 2004, in New Frontiers in
  Stellar Interferometry, Proceedings of SPIE Volume 5491. Edited by Wesley A.
  Traub. Bellingham, WA: The International Society for Optical Engineering,
  2004., p.528, ed. W.~A. {Traub}, 528--+

\bibitem[{{Groenewegen}(2008)}]{BAADE-WESSELINK}
{Groenewegen}, M.~A.~T. 2008, \aap, 488, 25

\bibitem[{{Keller} \& {Wood}(2006)}]{KELLER_2006}
{Keller}, S.~C. \& {Wood}, P.~R. 2006, \apj, 642, 834

\bibitem[{{Kervella} {et~al.}(2004){Kervella}, {Nardetto}, {Bersier},
  {Mourard}, \& {Coud{\'e} du Foresto}}]{KERVELLA_XSGR}
{Kervella}, P., {Nardetto}, N., {Bersier}, D., {Mourard}, D., \& {Coud{\'e} du
  Foresto}, V. 2004, \aap, 416, 941

\bibitem[{{Lafrasse} {et~al.}(2010){Lafrasse}, {Mella}, {Bonneau}, {Duvert},
  {Delfosse}, \& {Chelli}}]{Lafrasse_2010}
{Lafrasse}, S., {Mella}, G., {Bonneau}, D., {et~al.} 2010, VizieR Online Data
  Catalog, 2300, 0

\bibitem[{{Mathias} {et~al.}(2006){Mathias}, {Gillet}, {Fokin}, {Nardetto},
  {Kervella}, \& {Mourard}}]{Mathias_2006}
{Mathias}, P., {Gillet}, D., {Fokin}, A.~B., {et~al.} 2006, \aap, 457, 575

\bibitem[{{Petrov} {et~al.}(2007){Petrov}, {Malbet}, {Weigelt}, {Antonelli},
  {Beckmann}, {Bresson}, {Chelli}, {Dugu{\'e}}, {Duvert}, {Gennari},
  {Gl{\"u}ck}, {Kern}, {Lagarde}, {Le Coarer}, {Lisi}, {Millour}, {Perraut},
  {Puget}, {Rantakyr{\"o}}, {Robbe-Dubois}, {Roussel}, {Salinari}, {Tatulli},
  {Zins}, {Accardo}, {Acke}, {Agabi}, {Altariba}, {Arezki}, {Aristidi},
  {Baffa}, {Behrend}, {Bl{\"o}cker}, {Bonhomme}, {Busoni}, {Cassaing},
  {Clausse}, {Colin}, {Connot}, {Delboulb{\'e}}, {Domiciano de Souza},
  {Driebe}, {Feautrier}, {Ferruzzi}, {Forveille}, {Fossat}, {Foy},
  {Fraix-Burnet}, {Gallardo}, {Giani}, {Gil}, {Glentzlin}, {Heiden},
  {Heininger}, {Hernandez Utrera}, {Hofmann}, {Kamm}, {Kiekebusch}, {Kraus},
  {Le Contel}, {Le Contel}, {Lesourd}, {Lopez}, {Lopez}, {Magnard}, {Marconi},
  {Mars}, {Martinot-Lagarde}, {Mathias}, {M{\`e}ge}, {Monin}, {Mouillet},
  {Mourard}, {Nussbaum}, {Ohnaka}, {Pacheco}, {Perrier}, {Rabbia}, {Rebattu},
  {Reynaud}, {Richichi}, {Robini}, {Sacchettini}, {Schertl}, {Sch{\"o}ller},
  {Solscheid}, {Spang}, {Stee}, {Stefanini}, {Tallon}, {Tallon-Bosc}, {Tasso},
  {Testi}, {Vakili}, {von der L{\"u}he}, {Valtier}, {Vannier}, \&
  {Ventura}}]{Petrov_AMBER}
{Petrov}, R.~G., {Malbet}, F., {Weigelt}, G., {et~al.} 2007, \aap, 464, 1

\bibitem[{{Pietrzy{\'n}ski} {et~al.}(2011){Pietrzy{\'n}ski}, {Thompson},
  {Graczyk}, {Gieren}, {Pilecki}, {Udalski}, {Soszynski}, {Bono}, {Konorski},
  {Nardetto}, \& {Storm}}]{PIETRZYNSKI_2011}
{Pietrzy{\'n}ski}, G., {Thompson}, I.~B., {Graczyk}, D., {et~al.} 2011, \apjl,
  742, L20

\bibitem[{{Prada Moroni} {et~al.}(2012){Prada Moroni}, {Gennaro}, {Bono},
  {Pietrzy{\'n}ski}, {Gieren}, {Pilecki}, {Graczyk}, \&
  {Thompson}}]{PRADA_MORONI_2012}
{Prada Moroni}, P.~G., {Gennaro}, M., {Bono}, G., {et~al.} 2012, \apj, 749, 108

\bibitem[{{Sasselov} \& {Lester}(1990)}]{Saselov_Lester_1990}
{Sasselov}, D.~D. \& {Lester}, J.~B. 1990, \apj, 362, 333

\bibitem[{{Storm} {et~al.}(2011){Storm}, {Gieren}, {Fouqu{\'e}}, {Barnes},
  {Soszy{\'n}ski}, {Pietrzy{\'n}ski}, {Nardetto}, \& {Queloz}}]{STORM_2011}
{Storm}, J., {Gieren}, W., {Fouqu{\'e}}, P., {et~al.} 2011, \aap, 534, A95

\bibitem[{{Szabados}(1989)}]{SZABADOS_1989}
{Szabados}, L. 1989, Commmunications of the Konkoly Observatory Hungary, 94, 1

\bibitem[{{Szabados}(1990)}]{SZABADOS_1990}
{Szabados}, L. 1990, \mnras, 242, 285

\bibitem[{{Szabados}(2003)}]{SZABADOS_2003}
{Szabados}, L. 2003, Information Bulletin on Variable Stars, 5394, 1

\bibitem[{{Tatulli} {et~al.}(2007){Tatulli}, {Millour}, {Chelli}, {Duvert},
  {Acke}, {Hernandez Utrera}, {Hofmann}, {Kraus}, {Malbet}, {M{\`e}ge},
  {Petrov}, {Vannier}, {Zins}, {Antonelli}, {Beckmann}, {Bresson}, {Dugu{\'e}},
  {Gennari}, {Gl{\"u}ck}, {Kern}, {Lagarde}, {Le Coarer}, {Lisi}, {Perraut},
  {Puget}, {Rantakyr{\"o}}, {Robbe-Dubois}, {Roussel}, {Weigelt}, {Accardo},
  {Agabi}, {Altariba}, {Arezki}, {Aristidi}, {Baffa}, {Behrend}, {Bl{\"o}cker},
  {Bonhomme}, {Busoni}, {Cassaing}, {Clausse}, {Colin}, {Connot},
  {Delboulb{\'e}}, {Domiciano de Souza}, {Driebe}, {Feautrier}, {Ferruzzi},
  {Forveille}, {Fossat}, {Foy}, {Fraix-Burnet}, {Gallardo}, {Giani}, {Gil},
  {Glentzlin}, {Heiden}, {Heininger}, {Kamm}, {Kiekebusch}, {Le Contel}, {Le
  Contel}, {Lesourd}, {Lopez}, {Lopez}, {Magnard}, {Marconi}, {Mars},
  {Martinot-Lagarde}, {Mathias}, {Monin}, {Mouillet}, {Mourard}, {Nussbaum},
  {Ohnaka}, {Pacheco}, {Perrier}, {Rabbia}, {Rebattu}, {Reynaud}, {Richichi},
  {Robini}, {Sacchettini}, {Schertl}, {Sch{\"o}ller}, {Solscheid}, {Spang},
  {Stee}, {Stefanini}, {Tallon}, {Tallon-Bosc}, {Tasso}, {Testi}, {Vakili},
  {von der L{\"u}he}, {Valtier}, \& {Ventura}}]{Tatulli_AMBER}
{Tatulli}, E., {Millour}, F., {Chelli}, A., {et~al.} 2007, \aap, 464, 29

\end{thebibliography}

\end{document}